\documentclass[twocolumn,prl,preprintnumbers,nofootinbib,showpacs]{revtex4}

\usepackage{graphicx}
\usepackage{bm}
\usepackage{amssymb,amsmath}

\def\beqn{\begin{eqnarray}}
\def\eeqn{\end{eqnarray}}
\def\barr{\begin{array}}
\def\earr{\end{array}}
\def\btab{\begin{tabular}}
\def\etab{\end{tabular}}
\def\bite{\begin{itemize}}
\def\eite{\end{itemize}}
\def\bcen{\begin{center}}
\def\ecen{\end{center}}

\def\eq{\begin{equation}}
\def\ee{\end{equation}}
\def\nn{\nonumber}

\def\keldagger{k\hspace{-0.2cm}/}

\def\q2dagger{q_2\hspace{-0.35cm}/\;}

\begin{document}


\title{Nucleon EDM and rare decays of $\eta$ and $\eta$' mesons}
\author{Mikhail Gorchtein} 
\email{gorshtey@caltech.edu}
\affiliation{
Nuclear Theory Center and Physics Department, Indiana University,
Bloomington, IN 47403} 

\date{\today}

\begin{abstract}
I consider rare $CP$-violating decay modes
$\eta(\eta')\to\pi\pi$ and note that an interaction that leads to such decays 
would necessarily induce a nucleon electric dipole moment (EDM). 
The experimental limits for the corresponding branching ratios are quite soft. 
I relate these decay branching ratios to the value of the induced nucleon EDM 
and translate the experimental limits on the neutron EDM 
into much more stringent constraints on these decay rates: 
$\frac{\Gamma_{\eta\to\pi\pi}}{\Gamma_\eta}\lesssim3.5\times10^{-14}$ and 
$\frac{\Gamma_{\eta'\to\pi\pi}}{\Gamma_\eta'}\lesssim1.8\times10^{-17}$.
\end{abstract}


\maketitle


The first proposal of experimental search of $CP$ violation effects in atoms 
was made almost 40 years ago \cite{shapiro}. 
The modern advanced experimental techniques realized in the experiments 
on electron's electric dipole moment (EDM) are based on that idea and have 
the sensitivity of $d_e\sim10^{-26}\,e\,{\rm cm}$ \cite{electron_edmexp}.

Apart from the electron EDM, experimental searches for the EDM of the neutron 
are on-going.Current experimental limit on the value of the electric dipole 
moment (EDM) of the neutron is  
$d_n\lesssim 2.9\times10^{-26}e$ $cm$ \cite{neutron_edm}. From theoretical 
point of view, a non-zero EDM could imply non-zero values for the QCD 
$\theta$-term as the latter can induce an EDM \cite{theta_edm}. 

Recently, an experimental program has been under consideration at JLab Hall D, 
persuing the goal of a higher precision determination of the rare decay rates 
for processes $\eta(\eta')\to\pi^0\gamma\gamma$ and $\eta(\eta')\to\pi\pi$ 
within the GlueX experiment \cite{gluex}.
The latter decay channel explicitly violates 
the $CP$-conservation. It becomes clear if considering this decay in the rest 
frame of the initial $\eta$ meson. Since it is spinless, the resulting pair 
of pions can only be produced in the $S$-wave, but in this way the intrinsic 
parities of the initial and final state are opposite, while the $C$-parity is 
not affected. 

In this letter, I will address the limits on rare decays 
$\eta(\eta')\to\pi\pi$. I will show that provided such decays do go at a 
non-zero rate, this would induce an effective $CP$-violating $\eta NN$ coupling.
Consequently, the photon-nucleon coupling will obtain a $CP$-violating 
contribution due to virtual $\eta$-loops. Since such a coupling, the electric 
dipole moment (EDM) of the nucleon is tightly constrained by 
experimental observations, this imposes a very stringent constraint on the 
$\eta(\eta')\to\pi\pi$ decay branching ratios.

PDG quotes the following branching ratios for these decays \cite{PDG}:
\beqn
\frac{\Gamma(\eta\to\pi\pi)}{\Gamma_\eta^{full}}&<&
\left\{
\begin{array}{c}
3.3\times10^{-4},\;\pi^+\pi^-\\
4.3\times10^{-4},\;\pi^0\pi^0
\end{array}
\right. \nn\\
\frac{\Gamma(\eta'\to\pi\pi)}{\Gamma_{\eta'}^{full}}&<&2\%
\label{eq:explimits}
\eeqn
The masses and full widths of the two mesons are quoted in PDG \cite{PDG} as 
$m_\eta=547.75\pm0.12$ MeV, $\Gamma_\eta^{full}=1.29\pm0.07$ keV and
$m_{\eta'}=958.78\pm0.14$ MeV, $\Gamma_{\eta'}^{full}=0.202\pm0.016$ MeV, 
respectively.
\begin{figure}[th]
\includegraphics[height = 1cm]{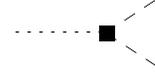}
\caption{$\eta(\eta')\to\pi\pi$ decay process. The solid square represents the 
$CP$-violating vertex.}
\label{fig:etapipi}
\end{figure}

The only form of the effective Lagrangian that can lead to the process shown in 
Fig.\ref{fig:etapipi} has the form
\beqn
{\cal{L}}=f_{\eta\pi\pi}m_\eta\eta\pi_a\pi_b\delta_{ab},
\eeqn
\noindent
with $f_{\eta\pi\pi}$ the corresponding coupling constant that is chosen to be 
dimesionsless, $m_\eta$ the mass of the $\eta$-meson, and the indices $a,b$ 
refer to isospin. Similar form holds for $\eta'$.

The correcponding decay widths can be calculated using the above Lagrangian:
\beqn
\Gamma_{\eta\to\pi\pi}=\frac{\sqrt{m_\eta^2-4m_\pi^2}}{16\pi}|f_{\eta\pi\pi}|^2
\label{eq:etapipi_decay_coupling}
\eeqn
\indent
The experimental limits of Eq. (\ref{eq:explimits}) can be translated into 
bounds on the coupling constants,
\beqn
f_{\eta\pi\pi}&\lesssim&2.3\times10^{-4}\nn\\
f_{\eta'\pi\pi}&\lesssim&1.5\times10^{-2}
\label{eq:etapipi_limit}
\eeqn
\indent
In turn, a $CP$-violating coupling of $\eta$'s to two pions will induce an 
effective $CP$-violating $\eta NN$ vertex
\beqn
{\cal{L}}_{\eta NN}^{CP}=\bar{g}_{\eta NN}\bar{N}N\eta
\eeqn
with the corresponding coupling $\bar{g}_{\eta NN}$. 
To estimate the value of this coupling, I will use heavy baryon ChPT 
formalism with pions, $\eta$ and $\rho$ mesons and the nucleon. I do not 
include the nucleon resonances and the $\Delta$ into this calculation. 
These resonances may be included in the effective field theory formalism 
as it was shown for the $\Delta$ \cite{vovchik} in principle, and the 
author expects sizeable contributions due to large $\pi N\Delta$ and 
$\eta N S_{11}(1535,1620)$ couplings. Perfectly aware of these shortcomings, 
I provide an order-of-magnitude estimate rather than the exact calculation. 

\begin{figure}[th]
\includegraphics[height = 2.5cm]{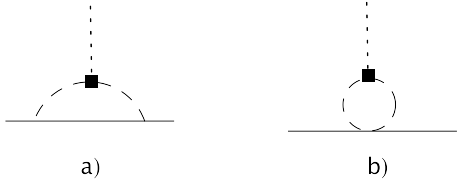}
\caption{Induced $CP$-violating $\eta(\eta')NN$ vertex. 
The solid square represents the $CP$-violating $\eta(\eta')\pi\pi$ vertex.}
\label{fig:etann}
\end{figure}

In Fig. \ref{fig:etann}, 
I display the two possible diagrams that can contribute to this vertex. 
However, only the diagram a) contribution is non-zero, while the ted-pole 
contribution (b graph) vanishes due to the isospin dependence.
The lowest order ChPT Lagrangian in the heavy baryon formalism is well known 
and I refer the reader to Ref. \cite{hbchpt} for the details. 
The calculation leads to the following relation:
\beqn
\bar{g}_{\eta NN}=-f_{\eta \pi\pi}m_\eta\left(\frac{g_{\pi NN}}{2M}\right)^2
\frac{m_\pi}{16\pi}
\left[\frac{2}{\pi}(\Delta+\ln\frac{\mu^2}{m_\pi^2})+1\right]
\eeqn
where $g_{\pi NN}=g_A\frac{M}{F_\pi}$ is the pion-nucleon coupling constant, 
$F_\pi$ being the pion decay constant. Similar result holds for $\eta'$.
The above loop calculation contains a 
divergence that is regularized by means of dimensional regularization 
with $\Delta=\frac{1}{\epsilon}-\gamma_E+\ln(4\pi)$ the usual $\bar{\rm MS}$ 
scheme subtraction, and $\mu$ is the renormalization scale that I will take 
to be equal to $\Lambda_{QCD}$. 
When performing the renormalization within the full 
underlying theory, the divergent part should cancel. I will use the remaining 
finite part to estimate the limits on the effective $CP$-violating $\eta NN$ 
coupling implied by the experimental limits on $\eta(\eta')$ decays,
\beqn
|\bar{g}_{\eta NN}|&\sim&f_{\eta \pi\pi}m_\eta
\left(\frac{g_{\pi NN}}{2M}\right)^2\frac{m_\pi}{16\pi}
\lesssim1.8\times10^{-5}\nn\\
|\bar{g}_{\eta'NN}|&\sim&f_{\eta'\pi\pi}m_{\eta'}
\left(\frac{g_{\pi NN}}{2M}\right)^2\frac{m_\pi}{16\pi}
\lesssim1.9\times10^{-3}
\label{eq:etann_limit}
\eeqn

\begin{figure}[th]
\includegraphics[height = 5.5cm]{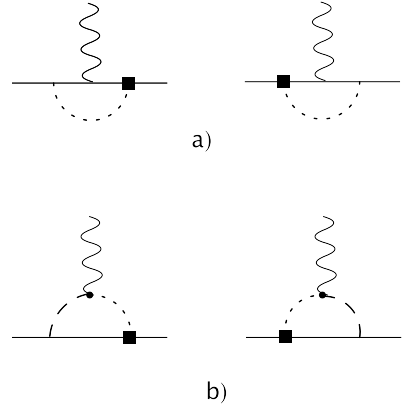}
\caption{Contribution to the nucleon EDM due to CP-violating $\eta(\eta')NN$ 
coupling. Contributions from $\eta$ loops (diagrams a) and 
$\eta\rho(\eta\omega)$ loops (diagrams b) are shown. 
Solid square represents the $CP$-violating $\eta NN$ vertex.}
\label{fig:etaedm}
\end{figure}

In the presence of the $CP$-violating coupling of $\eta$'s to the nucleon, 
the $\eta$'s can contribute to the 
nucleon EDM via virtual loops, as shown in Fig.\ref{fig:etaedm}. 

To estimate the contribution of such an interaction to the nucleon EDM, one 
can use effective field theory for $\eta$'s like for pions. 

The usual, $CP$-conserving $\eta NN$ coupling is given by
\beqn
{\cal{L}}_{\eta NN}=\frac{g_{\eta NN}}{2M}\bar{N}\keldagger\gamma_5N\eta
\eeqn
with $k$ the meson momentum. The couplings ${g}_{\eta NN}$ and 
${g}_{\eta' NN}$ are not very well known. The former can range 
between 0.5 and 1.5 \cite{lothar}, depending on the model which is used to 
calculate or extract it from the data. Furthermore, along with the PV coupling 
shown above, also the PS coupling ${g}_{\eta NN}\bar{N}\gamma_5N$ 
is allowed for the $\eta$'s unlike for pions. While they are equivalent for 
on-shell nucleons, inside the loop the use of the one or the other coupling 
will in general lead to different results. In the calculation, I will use the 
PV coupling and will assume that this PV coupling for $\eta$ and $\eta'$ 
are of order 1 and are roughly equal. 

Finally, the nucleon electromagnetic vertex with real photons is given by
\beqn
\Gamma^\mu(q)=e\left[e_N\gamma^\mu+\kappa_Ni\sigma^{\mu\alpha}
\frac{q_\alpha}{2M_N}
+\tilde{d}_Ni\gamma_5\sigma_{\mu\alpha}\frac{q_\alpha}{2M_N}\right]
\eeqn
where the dimensionless $\tilde{d}_N$ is the electric dipole moment (EDM) of 
the nucleon measured in units of the nuclear magneton $\frac{e}{2M_N}$, and the 
index $N=p,n$ indicates whether the nucleon is the proton or the neutron, 
respectively. 
Since the only direct experimental constraint on the value of the EDM is for 
the neutron, I will compute the induced neutron EDM in which case the photon 
couples to the anomalous magnetic moment of the neutron. 
The calculation is simplest to perform in HBChPT formulation.

To calculate the contribution of the vector meson loops shown in 
Fig. \ref{fig:etaedm},b, we need to define the effective Lagrangian for 
$\gamma\eta\rho(\gamma\eta\omega)$ vertex, as well as the vector meson 
coupling to the nucleon. These are given by \cite{lothar}:
\beqn
L_{\gamma\eta V}&=&\frac{e\lambda_V}{4m_\eta}\epsilon_{\mu\alpha\nu\beta}
F^{\mu\alpha}V^{\nu\beta}\eta,\nn\\
\Gamma^\mu_{VNN}&=&\bar{N}\left(g_v^V\gamma^\mu+g_t^Vi\sigma^{\mu\alpha}
\frac{q_\alpha}{2M}\right)\tau_VN
\eeqn
\noindent
with the electromagnetic field strength tensor 
$F^{\mu\alpha}=\partial^\mu A^\alpha-\partial^\alpha A^\mu$ and the vector 
meson field tensor $V^{\mu\alpha}=\partial^\mu V^\alpha-\partial^\alpha V^\mu$, 
respectively. The isospin factor $\tau_V$ is 1 for the $\omega$ and $\tau_3$ 
for $\rho^0$.

The $V\eta\gamma$ coupling constants values can be deduced from the 
$V\to\eta\gamma$ decay widths, 
\beqn
\Gamma_{V\to\eta\gamma} = \frac{\alpha_{em}}{24}
\frac{(m_V^2-m_\eta^2)^3}{m_V^3m_\eta^2}\lambda_V^2
\eeqn
and the $V\eta'\gamma$ couplings from the $\eta'\to V\gamma$ decay widths with
\beqn
\Gamma_{\eta'\to V\gamma} = \frac{\alpha_{em}}{8}
\frac{(m_V^2-m_\eta^2)^3}{m_\eta^5}{\lambda'}_V^2.
\eeqn

The empiric values of the parameters introduced above are summarized in 
Table \ref{tab}:
\begin{table}[h]
\vspace{1cm}
   \begin{tabular}{|l|l|l|l|l|l}
\hline
V  & $m_V$ (MeV) & $g_v^V$  &  $g_t^V/g_v^V$   &  $\lambda_V$ &  $\lambda'_V$ \\
\hline
$\rho$ & 775.8 & 2.4 & 6.1 & 0.9 & 1.18 \\
$\omega$ & 782.6 & 16 & 0. & 0.25 & 0.43 \\
\hline
   \end{tabular}
\caption{Parameters for the vector mesons.}
\label{tab}
\end{table}

I take the values of the $VNN$ couplings from \cite{lothar}, and obtain the 
$V\eta\gamma$ couplings from the corresponding decay widths given in \cite{PDG} 
as explained above.

With the use of these vertices, one obtains the estimate for the 
loop contribution shown in Fig. \ref{fig:etaedm}, $a$ and $b$:
\beqn
\tilde{d}^{a,\eta}_n&=&\kappa_n
\frac{g_{\eta NN}\bar{g}_{\eta NN}}{8\pi^2} I(m_\eta^2)
\label{eq:edmresult}\\
\tilde{d}^{b,V}_n&=&\lambda_V \tau_V
\frac{(g_v^V+g_t^V)\bar{g}_{\eta NN}}{8\pi^2} 
\frac{m_V^2 I(m_V^2) - m_\eta^2 I(m_\eta^2)}{m_V^2 - m_\eta^2}\nn,
\eeqn
where I used the notation
\beqn
I(m^2)&=&\Delta+\ln\frac{\mu^2}{M^2}-\frac{m^2}{2M^2}\ln\frac{m^2}{M^2}+2\nn\\
&-&\frac{\sqrt{m^2(4M^2-m^2)}}{M^2}
\left[\arctan\frac{2M^2-m^2}{m\sqrt{4M^2-m^2}}\right.\nn\\
&&\;\;\;\;\;\;\;\left.
+\arctan\frac{m}{\sqrt{4M^2-m^2}}\right]
\eeqn
\indent
In the above equation, $\mu$ stands for the regularization scale. In this case, 
the natural scale at which the divergencies cancel is at least of order of the 
mesonic masses in the loops, rather than $\Lambda_{QCD}$ as for the pion loops. 
For the numeric estimates, I will set this scale to the nucleon mass. 
The $\eta'$ contribution obtains by substituting the respective mass and 
couplings into Eq. (\ref{eq:edmresult}).

We can now proceed with numeric estimates for the induced neutron EDM. 
For this, I will require that every individual contribution to the neutron EDM 
does not exceed the experimental limit. Evaluating the expressions for each 
contribution, the most stringent limits for both $CP$-violating 
$\eta(\eta')NN$-couplings come from $\eta$ loops for the $\eta$, and $V\eta'$ 
loops for the $\eta'$. These constraints read
\beqn
|\bar{g}_{\eta NN}|&\lesssim&1.6\times10^{-10}\nn\\
|\bar{g}_{\eta'NN}|&\lesssim&5.8\times10^{-11}.
\label{eq:etann_new}
\eeqn

Of course, if a full EFT calculation would be possible, these bounds might 
shift either way due to enhancement or partial cancellation of different 
contributions, both among the calculated meson loop effects alone and with 
the contributions of the nucleon resonances that are not included in this 
analysis. If these latter effects will come with the same relative sign, 
the resulting limit on $\bar{g}_{\eta NN}$ will become 
stronger, then the results of Eq.(\ref{eq:etann_new}) will still hold. 
The only significant qualitative difference will arise if the effects that 
were not considered here would tend to cancel the loop contributions that I 
provided. However, such cancellations can only lead to significant changes 
in the estimates if they occur at 99\% or even more percent level, so that 
the bound on the $CP$-violation in $\eta$-decay is loosened by several orders 
of magnitude. 
An example of such cancellation is observed in the case of the magnetic 
polarizability of the nucleon where the large diamagnetic contribution from 
the pion loops are cancelled by a large paramagnetic contribution due to 
$N\to\Delta$ electromagnetic transition that comes with the opposite sign. 
As a result, the magnetic polarizability of the proton is about ten times 
smaller than its electric polarizability \cite{rcs}.

A precise cancellation of physically different contributions at a level of 1\% 
or even below that would indicate an existence of an unknown symmetry 
that prevents the mechanisms considered here from contributing to the EDM. 
The precision of our knowledge of the interactions of the $\eta$'s and vector 
mesons with the nucleons and nucleon resonances, both experimentally and 
theoretically, is far from the level that 
would allow to observe such a symmetry, if it is to exist at all.

Comparing now Eq.(\ref{eq:etann_new}) to the bounds derived from the 
experimental limits on the $CP$-violating $\eta$'s decays, 
Eq. (\ref{eq:etann_limit}), we find a discrepancy of five orders of magnitude 
for $|\bar{g}_{\eta NN}|$ and eight orders of magnitude for 
$|\bar{g}_{\eta'NN}|$. 
The recursive calculation leads to the EDM-induced constraint onto the 
$\eta\pi\pi$ coupling constants:
\beqn
f_{\eta\pi\pi}&\lesssim&2\times10^{-9}\nn\\
f_{\eta'\pi\pi}&\lesssim&4.3\times10^{-10}
\label{eq:etapipi_new}
\eeqn
\indent
Finally, recalling the relation between these couplings and the corresponding 
decay branching ratios obtained earlier in 
Eq.(\ref{eq:etapipi_decay_coupling}), 
it is possible to deduce the naturalness constraints onto these decay rates 
from the experimental limits on neutron EDM:
\beqn
\frac{\Gamma(\eta\to\pi\pi)}{\Gamma_\eta^{full}}&\lesssim&
3.5\times10^{-14},\nn\\
\frac{\Gamma(\eta'\to\pi\pi)}{\Gamma_{\eta'}^{full}}&\lesssim&1.8\times10^{-17}.
\eeqn
\indent
These limits are more stringent than the current experimental limits by 10 
orders of magnitude for $\eta$, and by 15 orders of magnitude for $\eta'$.

In summary, I considered the $\eta(\eta')\to\pi\pi$ decay channels 
that explicitly violate parity and time-reveresal conservation. I constructed a 
Lagrangian for such an interaction and derived the induced $CP$-violating 
coupling of $\eta$'s to the nucleon using heavy baryon ChPT.
If this coupling is non-zero, it generates a contribution to the 
electric dipole moment of the nucleon through virtual $\eta$, $\eta\rho^0$ and 
$\eta\omega$ loops. The tight experimental constraints on the neutron EDM 
lead to the conclusion that the current expreimental limits on 
branching ratios for the $\eta$s decaying to 
two pions are highly underconstrained. I provide the naturalness bounds on 
these branching ratios, 
$\frac{\Gamma_{\eta\to\pi\pi}}{\Gamma_\eta}\lesssim3.5\times10^{-14}$ and 
$\frac{\Gamma_{\eta'\to\pi\pi}}{\Gamma_\eta'}\lesssim1.8\times10^{-17}$.
These results indicate that a 
direct experimental search for the signal in these decay channels is not 
feasible in the near future.

\begin{acknowledgments}
The author is grateful to Charles Horowitz for useful discussions. 
The work was supported by the US NSF under grant PHY 0555232.
\end{acknowledgments}

\end{document}